# An Analytical Probabilistic Expression for Modeling Sum of Spatial-dependent Wind Power Output

Libao Shi, *Senior Member, IEEE,* Yang Pan, Yixin Ni

*Abstract*—Applying probability-related knowledge to accurately explore and exploit the inherent uncertainty of wind power output is one of the key issues that need to be solved urgently in the development of smart grid. This letter develops an analytical probabilistic expression for modeling sum of spatial-dependent wind farm power output through introducing unit impulse function, copulas, and Gaussian mixture model. A comparative Monte Carlo sampling study is given to illustrate the validity of the proposed model.

*Index Terms*—wind power, uncertainty, dependence, copula, Diagonal difference

## I. Introduction

When the large scale wind power is integrated into the power grid, the inherent uncertainty of wind power makes the operating mechanism of power system become more complicated and uncontrollable. So far, a lot of significant research achievements have claimed to represent the uncertainty of wind power output from the perspective of probability, involving statistical probability distribution models [1], and Gaussian mixture fitting models [2-3], and the probability distribution transformation technique [4] and copulas [5-6] are widely used to model the dependencies of multiple wind farm outputs. Furthermore, the copula based dependent discrete convolution [6] is formulated to handle the summation of dependent uncertainties (such as the total output of multiple wind farms) with varying degrees of success. However, to some extent, the existing approaches are mainly focused on numerical-analysis based idea, rarely discussed and studied from the viewpoint of inherent mechanism. In particular, some discrete terminals objectively existed in the probability density function (PDF) of wind farm output are not considered during modeling. In this letter, we develops an analytical probabilistic model to describe the uncertainty of sum of multiple wind farm outputs considering the spatial dependence, and the Monte Carlo simulation technique is employed to validate the effectiveness of the proposed model.

## II. Proposed Methodology

In this letter, we only discuss the dependence of multiple wind farms since there exists relatively weak correlation between wind farm and load [5].

### A. Analytical Probabilistic Model Considering Two Wind Farms

Here, we assume that there are two wind farms named as WF1 and WF2 in which the corresponding wind speeds obey Weibull and Gumbel statistical distributions, respectively. Based on our previous work [7], the PDFs (symbolled as $w_1(P_{wf1})$ and $w_2(P_{wf2})$) of the two wind farm outputs $P_{wf1}$ and $P_{wf2}$ can be derived respectively. Through integral operations, the corresponding CDFs (symbolled as $W_1(P_{wf1})$ and $W_2(P_{wf2})$) can be obtained, which are represented by the specified piece-wise functional relations with two distinct steps respectively. Assuming that the two wind farm outputs have a joint PDF $m(P_{wf1}, P_{wf2})$ and a joint CDF $M(P_{wf1}, P_{wf2})$, then the PDF $w(P_{wf})$ of the sum of the two wind farms outputs ($P_{wf}=P_{wf1}+P_{wf2}$) can be obtained by

$$w(P_{wf}) = \int_{-\infty}^{+\infty} m(P_{wf1}, P_{wf} - P_{wf1}) dP_{wf1} \quad (1)$$

When considering the dependence of $P_{wf1}$ and $P_{wf2}$, based on the Sklar's theorem, we have

$$m(P_{wf1}, P_{wf2}) = \frac{\partial M(P_{wf1}, P_{wf2})}{\partial P_{wf1} \partial P_{wf2}} = \frac{\partial C(W_1(P_{wf1}), W_2(P_{wf2}))}{\partial W_1(P_{wf1}) \partial W_2(P_{wf2})} \cdot \frac{\partial W_1(P_{wf1}) \partial W_2(P_{wf2})}{\partial P_{wf1} \partial P_{wf2}}$$
$$= c(W_1(P_{wf1}), W_2(P_{wf2})) \cdot w_1(P_{wf1}) \cdot w_2(P_{wf2}) \quad (2)$$

where the copula function $C(W_1(P_{wf1}), W_2(P_{wf2}))=M(P_{wf1}, P_{wf2})$ can be modeled by copula theory.

Substituting (2) into (1), then the $w(P_{wf})$ can be calculated as:

$$w(P_{wf}) = \int_{-\infty}^{+\infty} c(W_1(P_{wf1}), W_2(P_{wf} - P_{wf1})) \cdot w_1(P_{wf1}) \cdot w_2(P_{wf} - P_{wf1}) dP_{wf1} \quad (3)$$

Let $P_{wfr1}$ and $P_{wfr2}$ denote the rated outputs of WF1 and WF2, respectively. Considering that the $f_1(P_{wf1})$ or $f_2(P_{wf2})$ is modeled by unit impulse function when $P_{wf1}$ or $P_{wf2}=0$, $P_{wf1}=P_{wfr1}$ or $P_{wf2}=P_{wfr2}$ [7], the corresponding probability at corners of the closed region pertinent to the random wind farm output is not possible to be analytically solved by using (3). In this situation, we first try to work out the CDF of $P_{wf}$, $W(P_{wf})$, and then take derivative with respect to $P_{wf}$ to obtain $w(P_{wf})$. Here, we have

$$W(P_{wf}^*) = \Pr(P_{wf1} + P_{wf2} \leq P_{wf}^*) = W_I(P_{wf}^*) + W_{II}(P_{wf}^*) \quad (4)$$

In (4), the 1st item $W_I(P_{wf}^*)=\Pr(P_{wf1}>0 \& P_{wf2}>0 \& P_{wf1}+P_{wf2}\leq P_{wf}^*)$ denotes the probability within the region, which can be obtained by the integral of $m(P_{wf1}, P_{wf2})$. The 2nd item $W_{II}(P_{wf}^*)=\Pr[(P_{wf1}=0||P_{wf2}=0||P_{wf1}=P_{wfr1}||P_{wf2}=P_{wfr2}) \& (P_{wf1}+P_{wf2} \leq P_{wf}^*)]$ denotes the probability at corners of the region, which can be calculated by a diagonal difference method [8]. Therefore, the feasible region of $P_{wf}$ can be divided four parts as shown in Fig. 1. Region I represents $0<P_{wf}<P_{wfr1}$, in which the probability of $P_{wf1}=0$ and $P_{wf2}=0$ is involved. Region II represents $P_{wfr1}<P_{wf}<P_{wfr2}$, in which the probability of $P_{wf1}=0$, $P_{wf2}=0$, and $P_{wf1}=P_{wfr1}$ is involved. Region III represents $P_{wfr2}<P_{wf}<P_{wfr1}+P_{wfr2}$, in which the probability of $P_{wf1}=0$, $P_{wf2}=0$, $P_{wf1}=P_{wfr1}$, and $P_{wf2}=P_{wfr2}$ is involved. Besides, there exist four discrete terminals at $P_{wf}=0$, $P_{wf}=P_{wfr1}$, $P_{wf}=P_{wfr2}$, and $P_{wf}=P_{wfr1}+P_{wfr2}$, in which the corresponding impulse sizes need to be solved.

Due to space limitations, only the Region I is taken for example to illustrate the process of solving PDF of $P_{wf}$.

Regarding the probability within the Region I, it can be solved by using (3) directly. It should be noted that the Gaussian mixture model (GMM) as shown in (5) is applied to fit an item that is hard to be analytically expressed in theory.

$$\sum_{i=1}^{n} a_i \cdot \exp\left(-\frac{(P_{wf} - b_i)^2}{c_i}\right) \quad (5)$$

In Region I, there are only two corners, i.e. $P_{wf1}=P_{wfr1}$ and $P_{wf2}=P_{wfr2}$, which can be solved by using the diagonal difference method. The detailed CDF expression $W_{II}(P_{wf})$ is given in (5).

This work was supported by the National Natural Science Foundation of China (51777103)
Libao Shi, Yang Pan and Yixin Ni are with Graduate School at Shenzhen, Tsinghua University, Shenzhen 518055, China (email: shilb@sz.tsinghua.edu.cn).

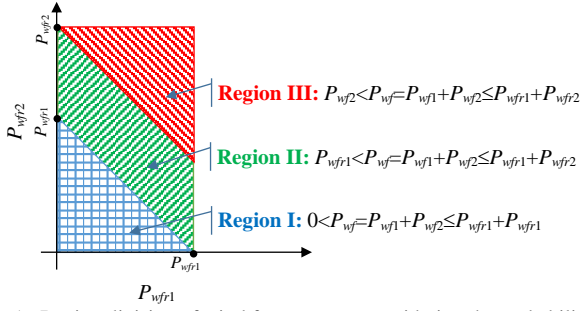

Fig. 1. Region division of wind farm outputs considering the probability at corners

$$W_{II}(P_{wf}) = M(0, P_{wf}) + M(P_{wf}, 0) - M(0,0) \quad (5)$$

Then, the derivative of $W_{II}(P_{wf})$ with respect to $P_{wf}$ can obtain the PDF $w_{II}(P_{wf})$ as described in (6).

$$w_{II}(P_{wf}) = \frac{\partial C(W_1(0), W_2(P_{wf}))}{\partial W_2} w_2(P_{wf}) + \frac{\partial C(W_1(P_{wf}), W_2(0))}{\partial W_1} w_1(P_{wf}) \quad (6)$$

So far, the PDF of total wind farm output in Region I has been solved analytically. Accordingly, the PDFs in Regions II and III can be derived in a similar way. The final analytical probabilistic expression of the PDF $w(P_{wf})$ is given in (7).

$$w(P_{wf}) = \begin{cases} \Phi_1(0)\delta(P_{wf}) & P_{wf} = 0 \\ \phi_1(P_{wf}) & 0 < P_{wf} < P_{wfr1} \\ \Phi_2(P_{wfr1})\delta(P_{wf} - P_{wfr1}) & P_{wf} = P_{wfr1} \\ \phi_2(P_{wf}) & P_{wfr1} < P_{wf} < P_{wfr2} \\ \Phi_3(P_{wfr2})\delta(P_{wf} - P_{wfr2}) & P_{wf} = P_{wfr2} \\ \phi_3(P_{wf}) & P_{wfr2} < P_{wf} < P_{wfr1} + P_{wfr2} \\ \Phi_4(P_{wfr1} + P_{wfr2})\delta(P_{wf} - P_{wfr1} - P_{wfr2}) & P_{wf} = P_{wfr1} + P_{wfr2} \end{cases} \quad (7)$$

where $\Phi_1(0)$, $\Phi_2(P_{wfr1})$, $\Phi_3(P_{wfr2})$, and $\Phi_4(P_{wfr1}+ P_{wfr2})$ denote the impulse sizes when $P_{wf}$ equals to 0, $P_{wfr1}$, $P_{wfr2}$, and $P_{wfr1}+P_{wfr2}$ respectively. The detailed expressions pertinent to $\Phi_1(0)$, $\Phi_2(P_{wfr1})$, $\Phi_3(P_{wfr2})$, $\Phi_4(P_{wfr1}+ P_{wfr2})$, $\phi_1(P_{wf})$, $\phi_2(P_{wf})$, and $\phi_3(P_{wf})$ are given as follows.

$$\Phi_1(0) = M(0,0)$$
$$\Phi_2(P_{wfr1}) = M(P_{wfr1}, 0) - M(P_{wfr1}^-, 0)$$
$$\Phi_3(P_{wfr2}) = M(0, P_{wfr2}) - M(0, P_{wfr2}^-) \quad (8)$$
$$\Phi_4(P_{wfr1} + P_{wfr2}) = 1 - M(P_{wfr1}^-, P_{wfr2}) - M(P_{wfr1}, P_{wfr2}^-) + M(P_{wfr1}^-, P_{wfr2}^-)$$

$$\phi_1(P_{wf}) = w_{II}(P_{wf}) + \int_0^{P_{wf}} m(P_{wf1}, P_{wf} - P_{wf1}) dP_{wf1} \quad (9)$$

$$\phi_2(P_{wf}) = \frac{\partial C(W_1(0), W_2(P_{wf}))}{\partial W_2} w_2(P_{wf}) + \\ [1 - \frac{\partial C(W_1(P_{wfr1}^-), W_2(P_{wf}-P_{wfr1}))}{\partial W_2}]w_2(P_{wf} - P_{wfr1}) + \\ \int_0^{P_{wfr1}^-} m(P_{wf1}, P_{wf} - P_{wf1}) dP_{wf1} \quad (10)$$

$$\phi_3(P_{wf}) = [1 - \frac{\partial C(W_1(P_{wf}-P_{wfr2}), W_2(P_{wfr2}^-))}{\partial W_1}]w_1(P_{wf} - P_{wfr2}) + \\ [1 - \frac{\partial C(W_1(P_{wfr1}^-), W_2(P_{wf}-P_{wfr1}))}{\partial W_2}]w_2(P_{wf} - P_{wfr1}) + \\ \int_{P_{wf}-P_{wfr2}}^{P_{wfr1}^-} m(P_{wf1}, P_{wf} - P_{wf1}) dP_{wf1} \quad (11)$$

### B. Analytical Probabilistic Model Expanding to N Wind Farms

The analytical probabilistic model proposed in this letter can be further expanded to the situation of more than two wind farms.

Taking $N$ wind farms for example, we assume that the outputs of $N$ wind farms are listed as $\{P_{wf1}, P_{wf2},...,P_{wfN}\}$. There are then $2*N$ corners: $\{P_{wf1}, P_{wf2},...,P_{wfi},...,P_{wfN}|P_{wfi}=0$ or $P_{wfi}=P_{wfri}\}$, where $P_{wfri}$ is the rated output of the $i$th wind farm. For the total output of $N$ wind farms, $P_{wf}=P_{wf1}+P_{wf2}+...+P_{wfN}$, there are $2^N$ critical points: 0, $P_{wfr1}, P_{wfr2},...,P_{wfrN}, P_{wfr1}+P_{wfr2},...,P_{wfr1}+P_{wfr2}+...+P_{wfrN}$. We arrange the $2^N$ critical points in ascending order, and then record them as $B_1, B_2,... B_{2^N}$. Accordingly, the PDF of the total output of the $N$ wind farms can be separately calculated in $2^N-1$ segments: $(B_1, B_2)$, $(B_2, B_3),...,(B_{2^N-1}, B_{2^N})$, with considering different corners as mentioned above.

## III. SIMULATION RESULTS

The Monte Carlo sampling (MCS) method is employed for comparative analysis to demonstrate the validity of the proposed model. Assuming that there are two wind farms that obey Weibull distribution with shape coefficient $k_w=2$, scale coefficient $\lambda_w=10$m/s, and Gumbel distribution with position coefficient $\mu_G=10$m/s, scale coefficient $\lambda_G=8$, respectively. The Gumbel copula with $\alpha_{Gumbel}=3.65$ is selected to model the dependence of the two wind farm outputs. Fig. 2 illustrates the results of proposed analytical probabilistic model and the MCS method with different sample sizes. It can be seen from Fig.2 that the curves obtained from the proposed model and the MCS method are perfectly matched, and the corresponding CPU time can be found in Table I.

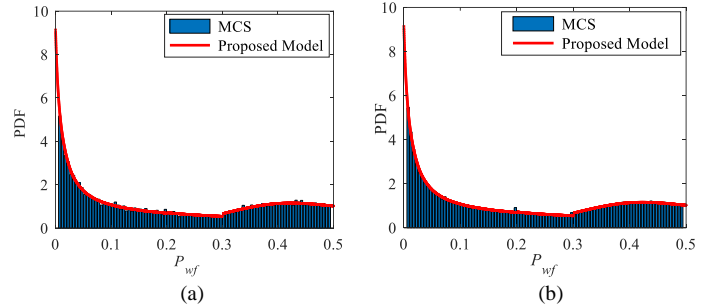

Fig. 2. Comparisons of the proposed model and MCS method. (a) Sample size=50000; (b) Sample size=200000.

TABLE I
CPU TIME APPLYING PROPOSED MODEL AND MCS METHOD

| MCS method | | Proposed model |
|---|---|---|
| Sample size=50000 | Sample size=200000 | |
| 2.6741s | 22.046s | 0.5208s |

## IV. CONCLUDING REMARKS

This letter proposed analytical probabilistic expression for modeling sum of spatial-dependent wind power output. Compared with the numerical simulation method, the proposed model can improve computational efficiency remarkably while keeping the accuracy. After expanding to $N$ wind farms, the proposed model bears potential and broad application prospects in stochastic optimization of smart grid considering renewable generation uncertainties.

## REFERENCES

[1] K. Tar "Some statistical characteristics of monthly average wind speed at various heights," *Renewable and Sustainable Energy Reviews*, vol. 12, no. 6, pp. 1712-1724, 2008.
[2] R. Singh, B. C. Pal, and R.A. Jabr, "Statistical representation of distribution system loads using Gaussian mixture model," *IEEE Trans. Power Syst.*, vol. 25, no. 1, pp. 29–37, Feb. 2010.
[3] Z. W. Wang, C. Shen, F. Liu, and F. Gao, "Analytical Expressions for Joint Distributions in Probabilistic Load Flow," *IEEE Trans. Power Syst.*, vol. 32, no. 3, pp. 2473-2474, 2017.
[4] Y. Chen, J. Wen, and S. Cheng, "Probabilistic load flow method based on Nataf transformation and Latin hypercube sampling," *IEEE Trans. Sustain. Energy*, vol. 4, no. 2, pp. 294–301, Apr. 2013.
[5] G. Papaefthymiou and D. Kurowicka, "Using copulas for modeling stochastic dependence in power system uncertainty analysis," *IEEE Trans. Power Syst.*, vol. 24, no. 1, pp. 40–49, 2009.
[6] N. Zhang, C. Q. Kang, C. Singh, and Q. Xia, "Copula Based Dependent Discrete Convolution for Power System Uncertainty Analysis," *IEEE Trans. Power Syst.*, vol. 31, no. 6, pp. 5204-5205, 2016.
[7] L. B. Shi, Z. X. Weng, L. Z. Yao, and Y. X. Ni, "An analytical solution for wind farm power output," *IEEE Trans. Power Syst.*, vol. 29, no. 6, pp. 3122-3123, 2014.
[8] S. Ferson, R. B. Nelson, J. Hajagos, et al. "Dependence in probabilistic modeling, Dempster-Shafer theory, and probability bounds analysis," Albuquerque: Sandia National Laboratories, 2004.